\documentstyle[12pt]{article}
\def\beq{\begin{equation}}
\def\eeq{\end{equation}}
\def\bea{\begin{eqnarray}}
\def\eea{\end{eqnarray}}
\def\bq{\begin{quote}}
\def\eq{\end{quote}}

\def\gappeq{\mathrel{\rlap {\raise.5ex\hbox{$>$}}
{\lower.5ex\hbox{$\sim$}}}}

\def\lappeq{\mathrel{\rlap{\raise.5ex\hbox{$<$}}
{\lower.5ex\hbox{$\sim$}}}}

\begin{document}
\topmargin -0.5cm
\oddsidemargin -0.3cm
\pagestyle{empty}
\begin{flushright} {CERN-TH.97-216}
\end{flushright}
\vspace*{5mm}
\begin{center} {\bf INDIRECT CP VIOLATION IN THE B-SYSTEM}\\   
\vspace*{1cm}  {\bf M.C. Ba\~{n}uls} \footnote{IFIC, Centro
Mixto Univ. Valencia - CSIC, E-46100 Burjasot (Valencia), Spain;
banuls@titan.ific.uv.es}
and  {\bf  J. Bernab\'eu} 
\footnote{Departamento de F\'{\i}sica Te\'orica, Univ. Valencia, E-46100
Burjasot (Valencia), Spain; bernabeu@evalvx.ific.uv.es} \\
\vspace*{1cm}
CERN - Geneva \\
\vspace*{2cm}   {\bf ABSTRACT} \\ \end{center}
\vspace*{5mm}
\noindent 
We show that, contrary to the flavour mixing amplitude $q/p$, both
$Re(\varepsilon)$ and $Im(\varepsilon)$ are observable quantities,
where
$\varepsilon$ is the phase-convention-independent CP mixing. 
We consider semileptonic $B_d$ decays from a CP tag and build appropriate time-dependent
asymmetries to separate out $Re(\varepsilon)$ and $Im(\varepsilon)$.
``Indirect" CP violation would have in $Im(\varepsilon)/(1+|\varepsilon|^2)$ its most prominent manifestation in the $B$-system, with expected values in the standard model ranging from $-0.37$ to $-0.18$. This quantity is controlled by a new observable ph
ase: the relative one between the CP-violating and CP-conserving parts of the effective hamiltonian.
For time-integrated rates we point out a $\Delta \Gamma \to \Sigma
\Gamma$ transmutation which operates in the perturbative CP mixing.
\vspace*{5cm}

\begin{flushleft} CERN-TH.97-216 
\end{flushleft}
\vfill\eject

\setcounter{page}{1}
\pagestyle{plain}

Since its discovery \cite{aaa} in 1964, CP violation has only been
seen in the $K^0 - \bar{K}^0$ system, in a few decay channels of the
long-lived kaon $K_L$ \cite{bb} and in a difference of decay rates
between $K^0$ and $\bar{K}^0$ \cite{cc}.  In the kaon system, the mechanism of
CP violation due to mixing of $K^0$ and $\bar{K}^0$ plays the most prominent
role and there is at present conflicting evidence on the existence of
``direct" CP violation in the decay amplitude.  Several experiments are
planned to measure CP violating parameters in $K$-physics.  CP violation can
be naturally described in the standard electroweak model as long as there are,
at least, three quark families \cite{dd}, whereby the elements of the quark
mixing matrix need not be relatively real. In the case of three families, the
standard model has a great deal of predictive power, as the complexity of the
quark mixing matrix is governed by a single weak phase. One of the most
important goals of particle physics is to determine precisely the elements of
the quark mixing matrix and test the standard model picture. The study of CP
violation in $B$ decays by means of dedicated experimental facilities in the
coming years can provide such an overdetermination of the parameters of the
quark mixing matrix.

In the case of the $B$ system, the standard model prospects \cite{ee} for
observation of CP violation due to \underline{flavour mixing} alone are quite
discouraging. The reason is that, to a good approximation, the flavour mixing
amplitude $q/p$ in the physical eigenstates of mass,
\bea
|B_1> & = & \frac{1}{\sqrt{|p|^2 + |q|^2}}
\left \{ p|B^0 > + q |\bar{B}^0 > \right \} \nonumber \\
|B_2> & = &  \frac{1}{\sqrt{|p|^2 + |q|^2}}
\left \{ p|B^0 > - q |\bar{B}^0 > \right \}
\label{I}
\eea
is just a pure phase. The parameter $q/p$ is phase-convention-dependent on
the definition of the CP-transformed states and thus its phase is not, by
itself, observable.  The best prospects \cite{ff} then make use of the
interplay between flavour mixing and decay.  The non-observability of the
flavour mixing phase is made apparent in the CP violating rate asymmetry, from
a \underline{flavour tag}, in the semileptonic decay $B^0 \to \ell \; \nu_l \;
X$:
\beq
a_{SL} \equiv \frac{N(\ell^+ \ell^+) - N(\ell^- \ell^-)}
{N(\ell^+ \ell^+) + N(\ell^- \ell^-)} = 
\frac{|p/q|^2 - |q/p|^2}{|p/q|^2 + |q/p|^2}
\label{II}
\eeq
To generate $|q/p| \not= 1$, one would need both $\Delta \Gamma_B \not= 0$
and a misalignment of the (complex) values of $\Gamma_{12}$ and $M_{12}$ in
the $B^0 - \bar{B}^0$ mass matrix.  All in all, the standard model $a_{SL}$ is
expected to be beyond the capabilities of the next experimental facilities,
although some prospects could appear for physics beyond the standard model
\cite{ggg}.

In this paper we show that, on the contrary, the semileptonic decay rate
asymmetry based on the \underline{CP-tag} has access to both $Re(\varepsilon)$
and $Im(\varepsilon)$, where $\varepsilon$ is the phase-convention-independent
parameter which governs \underline{CP-mixing} in the physical states
\bea
|B_1> = \frac{1}{\sqrt{1 + |\varepsilon |^2}} (|B_+ > + \varepsilon |B_- >)
\nonumber \\
|B_2> = \frac{1}{\sqrt{1 + |\varepsilon |^2}} (|B_- > + \varepsilon |B_+ >)
\label{III}
\eea
and $|B_{\pm} >$ are the CP eigenstates. In the literature it is common to
find the use of a different parameter, $\bar{\varepsilon}$, which controls the
mixing between the states $1/\sqrt{2} (|B^0 > \pm |\bar{B}^0 >)$.  These last
states are not the CP eigenstates unless we fix the phase to $|\bar{B}^0 >
\equiv \pm CP |{B}^0 >$.  As a consequence, the parameter $\bar{\varepsilon}$
changes under a phase redefinition. For an arbitrary phase convention, $|q/p|$
is connected to $Re (\varepsilon )$ as
\beq
\frac{2Re(\varepsilon )}{1 + |\varepsilon |^2} = 
\frac{1 - |q/p |^2}{1 + |q/p |^2}
\label{IV}
\eeq
The almost pure phase character of $q/p$ translates into a very small value of
$Re(\varepsilon )$. Experimentally \cite{hh}, one has $|Re(\varepsilon )| <
0.045$. From a flavour tag, the semileptonic decay of $B^0$ has no access to
$Im (\varepsilon )$. Both $Re(\varepsilon )$ and $Im(\varepsilon )$ are,
however, observable quantities.  It is of interest to illustrate by
model-independent methods how to separate out these two observables.

Let us assume that, at $t = 0$, the $B$-meson is prepared, in the quantum
mechanical sense, as a $|B_+ >$.  After this CP tag, the time evolution of 
$|B_+ >$ yields the probability amplitudes for the meson to behave as $|B_+ >, |B_- >$
\bea
|B_+ (t)> & = & \frac{1}{1-\varepsilon^2} \left \{[
e^{-i \lambda_1 t} - \varepsilon^2 e^{-i \lambda_2 t}] |B_+ > \right.
\nonumber \\
 & + & \left. \varepsilon [e^{-i \lambda_1 t} - e^{-i \lambda_2 t}]
|B_- > \right \}
\label{V}
\eea
where $ \lambda_j = m_j - i/2 \, \gamma_j \, \, (j = 1,2)$.  Equation (5) shows that
the CP mixing amplitude is linear in $\varepsilon$.  However, the survival
probability differs from the exponential decay law only by terms of order
$\varepsilon^2$, and the probability of becoming $|B_- >$ is of order 
$|\varepsilon|^2$.  More importantly, the observation of a final state $|\phi
>$ which is accessible from the two slits $|B_{\pm} >$ leads to an
interference pattern of order $\varepsilon$.  The corresponding decay rate can
be written as
\bea
|< \phi |B_+ (t) > |^2 & = & 
\frac{(1+|\varepsilon|^2)^2}{|1-\varepsilon^2|^2} \; 
e^{-\gamma_1 t}\; [a + b \; e^{-\Delta \Gamma t} \nonumber \\
& + & c \; e^{- \frac{\Delta \Gamma}{2} t} \; \cos (\Delta mt)
+ de^{- \frac{\Delta \Gamma}{2} t} \sin (\Delta mt)  ]
\label{VI}
\eea
where
\bea
a & = &\frac{1}{1+|\varepsilon|^2} \left [ \frac{|T_+|^2+|\varepsilon|^2|T_-|^2}{1+|\varepsilon|^2} + 2 Re \left ( \frac{\varepsilon}{1+|\varepsilon|^2} T^*_+ T_- \right ) \right ]
\nonumber \\
b & = & \frac{|\varepsilon|^2}{1+|\varepsilon|^2} \left [\frac{|T_-|^2+|\varepsilon|^2 |T_+|^2}{1+|\varepsilon|^2} + 2 Re \left ( \frac{\varepsilon^*}{1+|\varepsilon|^2} T^*_+ T_- \right ) \right ]  \nonumber \\
c & = &- 2\left [ Re \left [ \left ( \frac{\varepsilon}{1+|\varepsilon|^2}  + |\varepsilon|^2 \frac{\varepsilon^*}{1+|\varepsilon|^2}  \right )
T^*_+ T_- \right ] + Re \left (\frac{\varepsilon}{1+|\varepsilon|^2} \right )^2 |T_+|^2 + \left |  \frac{\varepsilon}{1+|\varepsilon|^2} \right |^2 |T_-|^2 \right ]
\nonumber \\
d & = & -2\left [ Im \left [\left (\frac{\varepsilon}{1+|\varepsilon|^2}  - |\varepsilon|^2 \frac{\varepsilon^*}{1+|\varepsilon|^2}  \right )
T^*_+ T_- \right ] + Im \left (\frac{\varepsilon}{1+|\varepsilon|^2} \right )^2 |T_+|^2 \right ]
\label{VII}
\eea
$T_{\pm} \equiv <\phi | B_{\pm} >$ are the decay amplitudes and 
$\Delta m - i \frac{\Delta \Gamma}{2} \equiv \lambda_2 - \lambda_1$.  For the
semileptonic decay $| \phi > = | \ell^- >$, the two amplitudes $T_{\pm}$ are
separately phase-convention-dependent, but the product $T_{+}^{*} T_-$ is not
and is real: due to the $\Delta B = \Delta Q_{\ell}$ rule, the charge of the
lepton selects the flavour and both $T_{+}$ and $T_{-}$ come from the same
flavour component.  For this semileptonic decay, we have
\bea
a & = & \frac{1}{2} |T|^2  \left [1 - 2 \frac{Re (\varepsilon )}{1+|\varepsilon|^2} \right ]\frac{1}{1+|\varepsilon|^2} \nonumber \\
b & = & \frac{1}{2} |T|^2  \left [1 - 2 \frac{Re (\varepsilon )}{1+|\varepsilon|^2} \right ]\frac{|\varepsilon|^2}{1+|\varepsilon|^2} \nonumber \\
c & = &|T|^2  \left [1 - 2 \frac{Re (\varepsilon )}{1+|\varepsilon|^2} \right ]\frac{ Re (\varepsilon )}{1+|\varepsilon|^2} \nonumber \\
d & = & |T|^2 \left [1 - 2 \frac{Re (\varepsilon )}{1+|\varepsilon|^2} \right ]\frac{ Im (\varepsilon )}{1+|\varepsilon|^2}
\label{VIII}
\eea
where $T \equiv < \ell^- | \bar{B}^0 >$ is the decay amplitude from the the
flavour state.  Equations (7) or (8) include an interference term, modulated
by $\sin (\Delta mt)$, proportional to $Im (\varepsilon )/(1+|\varepsilon|^2)$.

For the CP conjugate final state $| \bar{\phi} > = | \ell^+ >$ from the same
initial state, Eq. (6) has the same form with the coefficients of the
time-dependent terms replaced by
\bea
\bar{a} & = &  \frac{1}{2} |\bar{T}|^2  \left [1 + 2 \frac{Re (\varepsilon )}{1+|\varepsilon|^2} \right ]\frac{1}{1+|\varepsilon|^2}
\nonumber \\
\bar{b} & = &  \frac{1}{2} |\bar{T}|^2  \left [1 + 2 \frac{Re (\varepsilon )}{1+|\varepsilon|^2} \right ]\frac{|\varepsilon|^2}{1+|\varepsilon|^2} \nonumber \\
\bar{c} & = & - |\bar{T}|^2  \left [1 + 2 \frac{Re (\varepsilon )}{1+|\varepsilon|^2} \right ]\frac{ Re (\varepsilon )}{1+|\varepsilon|^2}
\nonumber \\
\bar{d} & = & - |\bar{T}|^2  \left [1 + 2 \frac{Re (\varepsilon )}{1+|\varepsilon|^2} \right ]\frac{ Im (\varepsilon )}{1+|\varepsilon|^2}
\label{IX}
\eea
where $\bar{T} \equiv < \ell^+ | B^0 >$.  CPT invariance imposes the equality
of probabilities $|\bar{T}|^2 = |T|^2$.  CP invariance is clearly violated if
\beq
a \neq \bar{a} \quad \mbox{or} \quad
b \neq \bar{b} \quad  \mbox{or} \quad
c \neq \bar{c} \quad \mbox{or} \quad
d \neq \bar{d} 
\label{X}
\eeq

To exhibit CP violation, we consider the corresponding CP asymmetry between Eqs. (8) and (9). To first order in $\varepsilon/(1+|\varepsilon|^2)$ we have
\bea
A^{CP}_{+} (t) & \equiv &
\frac{\Gamma [B_+ (t) \to \ell^+] - \Gamma [B_+ (t) \to \ell^-]}
{\Gamma [B_+ (t) \to \ell^+] + \Gamma [B_+ (t) \to \ell^-]} \nonumber \\
& = & 2 \frac{Re (\varepsilon )}{1+|\varepsilon|^2} [ 1 - e^{\frac{\Delta \Gamma}{2} t}
\cos (\Delta mt)] - 2 \frac{Im (\varepsilon )}{1+|\varepsilon|^2} e^{\frac{\Delta \Gamma }{2} t}
\sin (\Delta mt)
\label{XI}
\eea
The different time-dependence of the two terms on the right-hand side of Eq.
(11) allows a separation of both parts, $Re (\varepsilon )/(1+|\varepsilon|^2)$ and $Im(\varepsilon )/(1+|\varepsilon|^2)$, of the CP mixing parameter.

For a $t = 0$ preparation of the $B$-meson as $|B_- >$, the replacement
$B_+ \leftrightarrow B_-$ is accompanied by $\lambda_1 \leftrightarrow
\lambda_2$ in Eq. (5).  The decay rate can still be written as Eq. (6), but
\beq
a \to b, \quad b \to a, \quad c \to c, \quad d \to -d
\label{XII}
\eeq
with the simultaneous exchange $T_+ \leftrightarrow T_-$. The corresponding CP
violating asymmetry in semileptonic decay is then
\bea
A^{CP}_{-} (t) & \equiv & \frac{\Gamma [B_- (t) \to \ell^+ ] -
\Gamma [B_- (t) \to \ell^- ]}
{\Gamma [B_- (t) \to \ell^+ ] +\Gamma [B_- (t) \to \ell^- ]} \nonumber \\
& = & 2 \frac{Re (\varepsilon )}{(1+|\varepsilon|^2} [ 1 - e^{-\frac{\Delta \Gamma}{2} t}
\cos (\Delta mt)] + 2 \frac{Im (\varepsilon )}{1+|\varepsilon|^2} e^{-\frac{\Delta \Gamma}{2} t}
\sin (\Delta mt)
\label{XIII}
\eea
The analysis of Eq. (13) offers a complementary means to that of Eq. (11) for
separating out $Re (\varepsilon )/(1+|\varepsilon|^2)$ and $Im (\varepsilon )/(1+|\varepsilon|^2)$.

Let us now consider time-integrated rates to first order in $\varepsilon /(1+|\varepsilon|^2)$.
From Eqs. (6) and (8) it is straightforward to get for the semileptonic decay:
$$
\int^{\infty}_{0} \, dt |< \ell^- |B_+ (t) > |^2 = 
\frac{|T|^2}{2 \gamma_{2}} \;
\left \{ 1- 2 Re \left [\frac{\varepsilon}{1+|\varepsilon|^2} \frac{\Delta m + i \frac{\Delta \Gamma}{2}}
{\Delta m - i \frac{\Sigma \Gamma}{2}} \right ] \right \}
\eqno{(14.a)}
$$
$$
\int^{\infty}_{0} \, dt |< \ell^+ |B_+ (t) > |^2 =
\frac{|\bar{T}|^2}{2 \gamma_{2}} \;
\left \{ 1+ 2 Re \left [\frac{\varepsilon}{1+|\varepsilon|^2} \frac{\Delta m + i \frac{\Delta \Gamma}{2}}
{\Delta m - i \frac{\Sigma \Gamma}{2}} \right ] \right \}
\eqno{(14.b)}
$$
$$
\int^{\infty}_{0} \, dt |< \ell^- |B_- (t) > |^2 =
\frac{|T|^2}{2 \gamma_{1}} \;
\left \{ 1 - 2 Re \left [\frac{\varepsilon}{1+|\varepsilon|^2} \frac{\Delta m + i \frac{\Delta \Gamma}{2}}
{\Delta m + i \frac{\Sigma \Gamma}{2}} \right ] \right \}
\eqno{(14.c)}
$$
$$
\int^{\infty}_{0} \, dt |< \ell^+ |B_- (t) > |^2 =
\frac{|\bar{T}|^2}{2 \gamma_{1}} \;
\left \{ 1+ 2 Re \left [\frac{\varepsilon}{1+|\varepsilon|^2} \frac{\Delta m + i \frac{\Delta \Gamma}{2}}
{\Delta m + i \frac{\Sigma \Gamma}{2}} \right ] \right \}
\eqno{(14.d)}
$$
where $\Sigma \Gamma \equiv \gamma_1 + \gamma_2$. The simplicity and
interpretation of these results is amazing: in the time-integrated rates, the
effective CP-mixing [compare the signs for CP-conjugate decay modes (14.a) and
(14.b)] is not $\varepsilon/(1+|\varepsilon|^2)$ but that obtained by the recipe
$$
\frac{\varepsilon}{1+|\varepsilon|^2} \to  \frac{\varepsilon}{1+|\varepsilon|^2} \;
\frac{\Delta m + i \frac{\Delta \Gamma}{2}}
{\Delta m - i \frac{\Sigma \Gamma}{2}}
\eqno{(15)}
$$

In first order perturbation theory, the energy difference $\Delta m - i
\frac{\Delta \Gamma}{2}$ is expected to appear in the denominator of the
CP mixing parameter $\varepsilon/(1+|\varepsilon|^2)$. Equation (15) tells us that it is rather 
$\Delta m - i \frac{\Sigma \Gamma}{2}$ which is the relevant denominator.
Such a \underline{$\Delta \Gamma \to \Sigma \Gamma$ transmutation} was
noted \cite{jj} two decades ago in the context of parity violation by
neutral currents in muonic atoms \cite{kk}.  The appearance of the sum of
the widths is connected to the fact that the transitions from the two
admixed states $|B_+>$ and $|B_->$ are not resolved experimentally. The
occurrence of the sum of the widths implies that the maximum effect occurs
for $\Delta m = \frac{1}{2} \Sigma \Gamma $, a condition not far from
being valid for the $B_d$ system, where \cite{hh} $x_d = 
\frac{\Delta m}{\Gamma} = 0.73 \pm 0.05$.

Even if $\frac{\Delta \Gamma}{2 \Delta m}$ is expected to be very small in
the $B$-system \cite{ee}, of the order of $m^{2}_{b} / m^{2}_{t}$, the
comparable values of $\Delta m$ and $\frac{\Sigma \Gamma}{2}$ help the
objective of separating out $Re (\varepsilon )/(1+|\varepsilon|^2)$ and $Im (\varepsilon )/(1+|\varepsilon|^2)$
from the time-integrated rates given by Eqs. (14).  As emphasized above,
in the limit $\Delta \Gamma \to 0$, one expects $Re (\varepsilon ) \to 0$,
and the observation of CP mixing would be based on a non-vanishing value
of $Im (\varepsilon )/(1+|\varepsilon|^2)$.

In order to clarify the meaning of $Im(\varepsilon)/(1+|\varepsilon|^2)$, we obtain the expression for our phase-convention-independent $\varepsilon$ in terms of the matrix elements of the effective hamiltonian in the flavour basis, $H_{12}=<B^o|H|\bar{B}
^o>$. 
Both the dispersive part $M_{12}$ and the absorptive part $\Gamma_{12}$ of $H_{12}$ are phase-convention-dependent, so that only their relative phase, which determines $Re(\varepsilon)$, is physical and a manifestation of CP violation. 
There is, however, a third phase-convention-dependent matrix element which is involved in the connection between CP eigenstates and Flavour eigenstates: ${CP}_{12}=<B^o|CP|\bar{B}^o>$. The relative phase between $M_{12}$ and ${CP}_{12}$ is physical and a 
(new) measure of indirect CP violation. We obtain

$$
\varepsilon=\frac{Im(\Gamma_{12} {CP}^*_{12}) + 2 i Im (M_{12} {CP}^*_{12})}{ 2 Re(M_{12} {CP}^*_{12})-i Re(\Gamma_{12} {CP}^*_{12}) + \Delta m - \frac{i}{2} \Delta \Gamma}
\eqno{(16)}
$$
where $(\Delta m)^2 - \frac{1}{4}(\Delta \Gamma)^2=4|M_{12}|^2-|\Gamma_{12}|^2$. In the standard model, indirect CP violation in the B system is in fact dominated by $Im(\varepsilon)/(1+|\varepsilon|^2)$ and thus given by a CP phase in the dispersive part
 of the $B^o - \bar{B}^o$ mixing amplitude, relative to the phase convention ${CP}_{12}$.
The main contribution to this dispersive part comes \cite{lll} from the box diagram with the top quark running in the loop. Using the matrix elements $M_{12}$ and $\Gamma_{12}$ as given in Ref. \cite{lll}, compatible with ${CP}_{12}=-1$, we have calculate
d the values of $Im(\varepsilon)/(1+|\varepsilon|^2)$ in terms of the Wolfenstein \cite{mm} parametrization of the quark mixing matrix. In the limit of dominance of the intermediate top quark, for which $\varepsilon$ is purely imaginary, we get

$$
\frac{Im(\varepsilon)}{1+|\varepsilon|^2} \simeq \frac{Im(M_{12} CP^*_{12})}{\Delta m} \simeq -\frac{\eta (1- \rho)}{(1- \rho)^2 +\eta^2}
\eqno{(17)}
$$

Recent estimates \cite{nn} of $\eta$ and
$\rho$, constrained from existing measurements, give an appreciable
value for $Im (\varepsilon )/(1+|\varepsilon|^2)$ ranging from ${-0.37}$ to ${-0.18}$.

The reference phase $CP_{12}$ is given by the flavour mixing amplitude in the CP-conserving limit, $(q/p)_{CP}$:

$$
\frac{1-\varepsilon}{1+\varepsilon}=\frac{q}{p} \, CP_{12} =-\frac{(q/p)}{(q/p)_{CP}}
\eqno{(18)}
$$

Therefore, CP violation in a given system will be realized either by $|q/p| \neq 1$ or by a relative phase between the CP-violating and the CP-conserving flavour mixing amplitudes, or by both of them. Contrary to the $K$-system, this new observable phase
would have the most prominent role in $B$-physics, where $|q/p| \simeq 1$. 

To conclude, the rate asymmetries in semileptonic $B$ decays from a
CP-tag offer an illustration of the separate observable character of
$Re (\varepsilon )$ and $Im (\varepsilon )$, where $\varepsilon$ is our phase-convention-independent CP-mixing parameter.
Our study yields an additional result: for time-integrated
rates, the effective CP-mixing contains an ``energy difference"
denominator given by $\Delta m - i \; \frac{\Sigma \Gamma}{2}$, instead
of $\Delta m - i \; \frac{\Delta \Gamma}{2}$. 
This allows the extraction of $Im(\varepsilon)/(1+|\varepsilon|^2)$, whose values in the standard model are expected to range from $-0.37$ to $-0.18$. The new phase discussed in this paper can be understood as the relative phase between the CP-violating a
nd CP-conserving parts of the effective hamiltonian.

\vspace{1cm}

{\bf Acknowledgements}\\

The authors are pleased to thank F.J. Botella, G.C. Branco, M.B. Gavela,
M. Gronau, P. Hern\'andez, T. Nakada, A. Pich, M. Rebelo and E. Roulet for their
comments and interesting discussions on the topic of this paper, and the
CERN Theory Division for its hospitality.  This research was supported
by CICYT, Spain, under grant No. AEN-96/1718.

\end{document}